\documentclass[reprint,nofootinbib,superscriptaddress,showpacs]{revtex4-1}
\usepackage[caption=false]{subfig}
\usepackage{hyperref}
\usepackage{t1enc}
\usepackage[latin2]{inputenc}
\usepackage{color}

\usepackage{amsmath,amsfonts,amssymb}
\usepackage{MnSymbol}
\usepackage{bbm}

\usepackage{epsfig}

\newcommand{\ve}[1]{\mathbf{#1}}

\newcommand{\avr}[1]{\langle #1\rangle}
\definecolor{!R}{rgb}{1,0,0}
\definecolor{!G}{rgb}{0,1,0}
\definecolor{!B}{rgb}{0,0,1}

\newcommand{\intlim}[3]{\int\limits_{#2}^{#3}\!\!\mathrm{d}#1}
\newcommand{\intpos}[1]{\int\limits_{#1}^+\!\!}
\newcommand{\Int}[1]{\int\limits_{#1}\!\!}
\newcommand{\re}{\mathrm{Re}}
\newcommand{\im}{\mathrm{Im}}

\newcommand{\kelphi}[2]{\varphi_{#1}^{(#2)}}
\newcommand{\kelphid}[2]{(\varphi_{#1}^\dagger)^{(#2)}}

\begin{document}
\title{Shear viscosity over entropy density ratio with extended quasi-particles}
\author{M. Horv\'ath}
\email[e-mail: ]{horvath.miklos@wigner.mta.hu}
\affiliation{Department of Theoretical Physics, Wigner Research Centre for Physics, Institute for Particle and Nuclear Physics, Budapest, Hungary}
\author{A. Jakov\'ac}
\affiliation{Institute of Physics, E\"otv\"os University, Budapest, Hungary}

\begin{abstract}
We consider an effective field theory description of beyond-quasi-particle excitations aiming to associate the transport properties of the system with the spectral density of states. Tuning various properties of the many-particle correlations, we investigate how the robust microscopic features are translated into the macroscopic observables like shear viscosity and entropy density. The liquid-gas crossover is analysed using several examples. A thermal constraint on the fluidity measure $\eta/s$ is discussed. 
\end{abstract}
\maketitle

\section{Introduction}
\begin{figure}
\centering
  \includegraphics[width=0.9\linewidth]{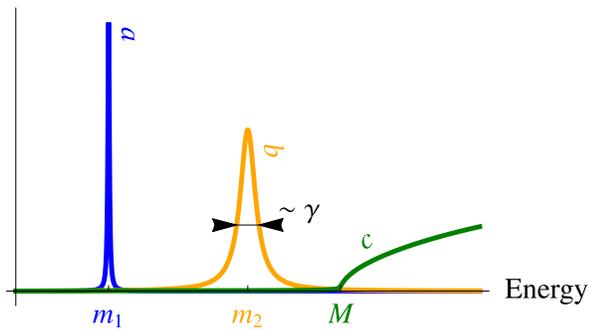}
  \caption{Robust features of a generic spectral function (color online). \textit{QP-behavior}: practically infinitely long lifetime (blue, $a$), \textit{Broad peak} with lifetime $\sim 1/\gamma$ (orange, $b$), \textit{Continuum of multi-particle states} with threshold $M$ (green, $c$)}
\label{fig:spectralfunctions}
\end{figure}

Thermodynamic and transport properties of physical systems are of great interest to theoretical investigations, since those are essential to explore the phase diagram of a given material, and to characterize the properties of its collective behaviour. These measurable quantities also give a basis for the comparison of the theoretical predictions to the physical reality. Despite the diversity of models, concepts like conductivity, viscosity, densities of energy and entropy etc.\ allow us to phenomenologically access a wider range of physical systems from cold atomic gases through fluids at room temperature to the hot and dense matter created in heavy-ion collisions. These macroscopic observables usually are in a very complicated relationship with the microscopic quantities (i.e.\ the fundamental degrees of freedom) of a given theory. There are numerous examples in the literature illustrating this elaborate issue, see for example Refs.~\cite{thermo_gluondamping, thermo_gluonplasma, thermo_quasipart0, thermo_quasipart1, thermo_quasipart2, thermo_quasipart3, thermo_selfcons1, thermo_selfcons2, thermo_QCDlattice} for the analysis concerning thermodynamical quantities. Furthermore see Refs.~\cite{Jeon, transport_YMfrg, transport_NJL3, transport_2PI1, transport_largeN1, transport_largeN2, transport_eff2, transport_SU3lattice, transport_2PI2, transport_eff1} for transport coefficients obtained from quantum field theory (QFT), functional renormalization group (FRG) or lattice calculations, and see Refs.~\cite{transport_QMD, transport_NJL4, transport_eff3, transport_eff4, transport_eff5} for kinetic theory or quasi-particle (QP) approaches. \\
Interestingly enough, the ratio of the shear viscosity $\eta$ to the entropy density $s$ has qualitatively the same temperature dependence in several systems, showing in general a fluid-like behaviour. Near to the critical endpoint of the liquid-gas phase-transition the fluidity measure $\eta/s$ achieves its minimal value \cite{transport_fermigas, transport_graphene, transport_exp1}, indicating that these materials are most fluent near to their critical state. \\
Our goal in this paper is to analyse the transport coefficient $\eta$ and the thermodynamic quantities in the framework of an effective field theory. We quantify how the robust properties of microscopically meaningful quantities relate to the qualitative behaviour of macroscopic observables. We use the spectral density of states or spectral function for this purpose, as it is meaningful even on the level of the fundamental theory. The spectral function $\rho_{x,y}$ is the response of the theory at the space-time point $y$ to a small, local perturbation occurred at $x$. In the momentum space, it characterizes the density of the quantum states in the energy $\omega$ if all other quantum numbers (including the momentum $\ve{p}$) kept fixed. Roughly speaking, $\rho_{\omega,\ve{p}}\mathrm{d}\omega$ quantifies the probability of the creation of an excitation with momentum $\ve{p}$ and energy within the interval ${[\omega,\omega+\mathrm{d}\omega]}$. A physically important characterization of $\rho_{\omega,\ve{p}}$ is whether it has a narrow-peak structure or not (see Fig.~\ref{fig:spectralfunctions}). If so, the behaviour of the system is dominated by (quasi-)particles, with inverse lifetime proportional to the half-width of the peak and with dispersion relation $\omega(\ve{p})$ determined by the position of the peak. Kinetic description and perturbation theory work usually well in this case. On the other hand for wide peak(s) or in the presence of a relevant continuum contribution, the situation is more intricate. The continuum contribution to $\rho_{\omega,\ve{p}}$ signals that multi-particle states are significant. Such spectra are produced by non-perturbative methods, for example the resummation of the infrared (IR) contributions of the perturbation theory \cite{BNresum1, BNresum2, BNresum3} or FRG calculations. Typically, the phenomenology of such systems cannot be described in terms of conventional quasi-particles with long lifetime. \\
The structure of this paper is the following. We summarize first the concept of quasi-particles and its limitations in effective modelling in Section \ref{exQP}. We introduce thermodynamic notions through the energy-momentum tensor in Section \ref{thermo}. The issue of thermodynamic consistency is briefly discussed. In Section \ref{linRes}. the transport coefficients in linear response are elaborated using Kubo's formula. After a short discussion on the lower bound of the ratio $\eta/s$ in the extended quasi-particle picture in Section \ref{lowerB}., we turn to analyse physically motivated examples in Section \ref{examples}.

\section{Extended quasi-particles}\label{exQP}
We concentrate to the transition between hydrodynamical and kinetic regimes. For this purpose, generalization of the notions of the QP-description is needed. From phenomenological point of view, quasi-particles are objects with infinite (or with very long) lifetime, usually well-localized in space. Resonances and other short-living yet particle-like entities are also often referred to as quasi-particles, confusingly.\\
From the side of QFT, particles are the asymptotic states of the theory in question. This definition, however, does not cover finite lifetime particle-like intermediate states often appearing in particle physics experiments. In effective modelling, one possibility is to associate a new field degree of freedom to every observed particle-like object. But non-physical symmetries could be generated via this resonance--field correspondence, it is not obvious how to avoid the double-counting of thermodynamic degrees of freedom \cite{SpectrFuncThermo}. \\
Finite lifetime bound-states and resonances are more natural to appear via interaction among some elementary fields. It is very unlikely though to guess those fundamental structures when constructing an effective theory, due to the lack of basic understanding, the reason we needed effective description in the first place. When the width is large, we must not rely on perturbative treatment any more: in this case the effective field theory approach helps to re-define the fundamental structures. \\
Let us consider a scalar (spin-0) operator $\varphi$ bearing all the physical degrees of freedom we are interested in. We call $\varphi$ an \textit{extended quasi-particle (EQP)} if its equation of motion is linear in $\varphi$. An equivalent statement is that the action is a quadratic functional of $\varphi$: $S[\varphi]=\frac{1}{2}\int_x\int_y\varphi_x\mathcal{K}_{x-y}\varphi_y$, and therefore the equation of motion (EoM) reads as $\int_y\mathcal{K}_{x-y}\varphi_y=0$. If so, all correlation functions are determined by the single two-point function $\rho_x=\avr{[\varphi_x,\varphi_0]}$, the so-called \textit{spectral function}, through Wick's theorem and causality. \\
In other words we use wave-packet-like modes instead of plane waves. The idea of using suitable basis of quantization, chosen to the actual problem, is widely used e.g.\ in solid state physics, like Cooper-pairs in superconductivity or atomic orbits and Wannier-functions in the description of crystals \cite{wannier1}. Choosing the appropriate degrees of freedom, the theory of strongly interacting elementary objects can become a weakly (or non-) interacting theory of composite ones. \\
The action of the EQP-description is non-local in the sense that the two field operators are inserted in different space-time points. There are several known examples in the literature for theories with non-local quadratic action, including pure gauge theories, low-energy effective theories of particles etc.\, see for example Refs.~\cite{nonlocal_scalar1, nonlocal_scalar2, nonlocal_resummation, nonlocal_gauge1, nonlocal_gauge2, nonlocal_condmat1}. \\
We view the non-local EQP-action as the leading order (or the relevant part) of an IR-resummed theory. Our goal is to describe the physics near to a critical point where second order phase transition occurs. We assume that the relevant field operator remains unchanged, but, since the long-range correlations may also play a major role, we allow the appearance of derivative terms in arbitrary order. These are the physical criteria cumulated in the non-local quadratic action. The quasi-particle nature is reflected in the linearity of the EoM, i.e.\ any linear combination of solutions also satisfies the field equation.\\
We stress here the main advantage of the quadratic nature of the description, the integrability, which allows us to calculate thermodynamic observables using two-point functions. Also in the case of transport coefficients where higher correlators are needed, the knowledge of two-point functions is sufficient for the linear response calculations since $\avr{\varphi\varphi\varphi\varphi} \sim \sum\avr{\varphi\varphi}\avr{\varphi\varphi}$, where the summation runs over all the possible pairings of the field operators $\varphi$.

\section{Equilibrium thermodynamics}\label{thermo}
Thermodynamic quantities (if no conserved charges are present) can originate from the energy density $\varepsilon$ or from the free energy density $f$. In both cases the averaging is performed over spatially translational invariant field configurations. Despite the lack of a well-defined canonical formalism, in non-local theories, $e^{-\beta T^{00}}$ serves as the usual Boltzmann statistical operator. With the time-evolution operator $e^{itT^{00}}$ the Kubo-Martin-Schwinger-relation holds, see Appendix \ref{propagators}, \ref{enmomTensor} and Ref.~\cite{SpectrFuncThermo} for further details. \\
Due to the quadratic form of the action, thermodynamic quantities can be expressed using the spectral function $\rho(\omega, |\ve{p}|)$ and the Fourier-transformed kernel $K(\omega, |\ve{p}|)$, see Ref.~\cite{SpectrFuncThermo}:
\begin{eqnarray}
f &=& -\intpos{p}\intlim{\tilde{\omega}}{}{\omega}\frac{\partial K(\tilde{\omega},|\ve{p}|)}{\partial\tilde{\omega}}\rho(\tilde{\omega},|\ve{p}|) n(\omega/T)  =-P,\label{pressure0} \\
\varepsilon &=& \intpos{p}\omega\frac{\partial K(\omega,|\ve{p}|)}{\partial\omega}\rho(\omega,|\ve{p}|) n(\omega/T). \label{energy0}
\end{eqnarray}
Here we used the notation $\intpos{p}\equiv\int\!\!\frac{\mathrm{d}^3\ve{p}}{(2\pi)^3}\int\limits_{0}^{\infty}\!\!\frac{\mathrm{d}\omega}{2\pi}$ for phase-space integration with respect to the four-momentum $p=(\omega,\ve{p})$, restricted to positive frequencies. Note, that $\rho$ and $K$ are not independent: $\rho=-\mathrm{Im}G(\omega+i0^+,\ve{p})$ and $K(\omega,\ve{p})=\mathrm{Re} G^{-1}(\omega+i0^+,\ve{p})$ with $G(\omega,\ve{p})=\intlim{\tilde{\omega}}{-\infty}{\infty}\frac{\rho(\tilde{\omega},\ve{p})}{\omega-\tilde{\omega}}$.

\subsection{Thermodynamic consistency}\label{thermoConsist}
We wish to also include systems with temperature-dependent parameters into our description. The consistency of Eq.~(\ref{pressure0}) and (\ref{energy0}) is fulfilled, however, only if $\rho$ and $K$ are temperature-independent. It means that the relations $s=\frac{\partial P}{\partial T}$ and $sT=\varepsilon+P$ hold, therefore $\varepsilon=T^2\frac{\partial (P/T)}{\partial T}$. \\
To overcome this issue and also keeping the simplicity of the EQP-picture we let $\phi:=\avr{\varphi}$ be non-zero, homogeneous and temperature-dependent. This is equivalent with a non-trivial, temperature-dependent ''bag constant'', see Ref.~\cite{thermo_quasipart0}. The correlators are shifted, thus $\varepsilon =\varepsilon_{\phi\equiv 0} +B$, $P=P_{\phi\equiv 0}-B$, with the temperature-dependent quantity $B$ (referring to the ''background''). This procedure leaves the entropy formula $s=\frac{\varepsilon+P}{T}$ unchanged (for further details see Appendix \ref{sourceTerm}). That is, the thermodynamic consistency is fulfilled using the same entropy formula, with temperature-dependent spectral function ${~\rho(\omega,\ve{p},\{m_i(T)\})}$. The background field is not arbitrary, its effect precisely cancels the extra terms coming from the temperature-dependent parameters $m_i$: ${\frac{\partial B}{\partial T}=\sum_i\frac{\partial m_i}{\partial T}\frac{\partial P_{\phi\equiv 0}(T,\{m_i(T)\})}{\partial m_i}}$.

\subsection{Microcausality}\label{mCausality}
Microcausality (or also often referred as locality) means that there is no correlation between two space-time points separated by a space-like interval. Since in our description all measurable quantity can be expressed by the spectral function, $\rho(x-y)\equiv 0$ is required for spatially separated space-time points $x$ and $y$. \\
In case of self-consistent approaches or perturbative calculations, microcausality is guaranteed by construction, as it is originated from the non-interacting theory and the space-time-local interaction vertices. In effective theories, this is not necessarily true. In order to guarantee microcausality, we choose the Fourier-transform of $\rho$ as ${\rho(\omega,\ve{p})=\theta(\omega^2-\ve{p}^2)\mathrm{sign}(\omega)\overline{\rho}(\omega^2-\ve{p}^2)}$, which simplifies Eq.~(\ref{pressure0}), (\ref{energy0}):
\begin{eqnarray}
P &=&\intlim{p}{0}{\infty}\frac{\partial K}{\partial p}\overline{\rho}(p)T^4 \chi_P(p/T), \label{pressure1} \\
\varepsilon &=&\intlim{p}{0}{\infty}\frac{\partial K}{\partial p}\overline{\rho}(p)T^4 \chi_\varepsilon(p/T), \label{energy1}
\end{eqnarray}
with the notation $p^2=\omega^2-\ve{p}^2$. The thermodynamic weight-functions are:
\begin{eqnarray}
\chi_P(x)&=& \frac{x^3}{4\pi^3}\intlim{y}{1}{\infty}y\sqrt{y^2-1}\cdot n(xy) \approx \frac{x^2}{4\pi^3}K_2(x), \\
\chi_\varepsilon(x)&=& \frac{x^4}{4\pi^3}\intlim{y}{1}{\infty}y^2\sqrt{y^2-1}\cdot n(xy) \approx \nonumber \\
&\approx & \frac{x^3}{4\pi^3}K_1(x)+\frac{3x^2}{4\pi^3}K_2(x),
\end{eqnarray} 
where $n$ is the Bose--Einstein distribution, $~{K_1,\, K_2,\, \dots}$ are modified Bessel functions appearing in the limit of the Boltzmannian approximation, when $n(x)\approx e^{-x}$. $T^4\chi_P(x)$ and $T^4\chi_\varepsilon(x)$ are the densities of pressure and energy of an ideal gas, respectively, with temperature $T$ and particle mass $xT$. It is apparent that the combination $K'(p)\overline{\rho}(p)$ acts as a mass-distribution (i.e.\ normalizable\footnote{$\int_0^\infty\!\!\mathrm{d}pK'(p)\overline{\rho}(p)$ has to be finite because of the existence of the high-temperature Stefan-Boltzmann-limit of the thermodynamics.}), therefore our quasi-particle description of the thermal observables can be interpreted as a mass-distributed ideal gas \cite{thermo_quasipart1, thermo_quasipart2}.\\
Note here, that the temperature-dependence of $\rho$ can break the manifest Lorentz-covariance through the temperature-dependent parameters, which are thought to be measured in the frame assigned to the heat bath.

\section{Shear viscosity in linear response}\label{linRes}
Hydrodynamics describes the collective motion of fluids with given material properties, based on the analysis of the energy-momentum conservation during the motion. The relaxation time of the system after a macroscopic perturbation is measured by the hydrodynamic transport coefficients. In case of a given transverse wave with wavenumber $\ve{k}$ perpendicular to the local flow velocity $\ve{v}$, its relaxation to the equilibrium configuration is controlled by $\eta/s$. Expressed with the energy-momentum tensor: ${\pi_\perp^{\mu\nu}=\pi_{\perp,0}^{\mu\nu}+\delta \pi_\perp^{\mu\nu}}$, the fluctuation part decays as ${\delta \pi_\perp^{\mu\nu}(t)=e^{-\frac{\eta}{s}\frac{\ve{k}^2t}{T}+i(\ve{k}\cdot\ve{r}-\omega t)}\delta \pi_\perp^{\mu\nu}(0)}$, where $T$ is the local temperature and $\omega=c_s|\ve{k}|$ with the sound velocity $c_s$, see Ref.~\cite{hydroLectures} for further details. \\
Transport coefficients can also be interpreted from the kinetic theory point of view. The shear viscosity $\eta$ is the diffusion coefficient of momentum transfer perpendicular to the local velocity of the fluid. In case of a gas of particles $\eta \sim v\lambda\rho$ with $v$ being the root mean square particle velocity, $\rho$ is its mass density and $\lambda$ is the mean free path of gas particles. Typically speaking, $\eta$ is large (compared to some internal scale) in gases (or in fluids where kinetic description is acceptable) compared to ordinary liquids. From the kinetic point of view, it means that the mean free path is significantly smaller in liquids. \\
A possible way to connect these two regimes is to define the transport coefficients in the linear response approximation. This allows us to go beyond the quasi-particle picture used in the kinetic theory and discuss the transport properties translated to those of a continuous medium represented by its energy-momentum tensor. Let us take a small perturbation in the action: ${~\delta S=\int_x h_xA_x}$, where $A$ is a measurable quantity (a Hermitian operator) and $h$ is a scalar function. The change in the expectation value of $B$ can be expressed up to first-order in $h$ as ${\delta\avr{B_x}=\int_yi\mathcal{G}^{ra}_{BA}(x-y)h_y}$. Kubo's formula characterizes the response function, supposing the system relaxes to thermal equilibrium in which it was before the perturbation occurred: ${i\mathcal{G}^{ra}_{BA}(x-y)=\theta_{x^0-y^0}\avr{[B_x,A_y]}=\theta_{x^0-y^0}\rho_{BA}(x-y)}$, where $\avr{.}$ refers to averaging over configurations in thermal equilibrium. \\ 
We intend to get the transport coefficients using a field theory framework. In case of the shear viscosity we are interested in the linear response to a small perturbation in the energy-momentum tensor $T^{\mu\nu}$. For the response function we need the spectral function $\rho_{TT}$, which we give in Appendix \ref{visco} in details. In the limit of long-wavelength (i.e.\ hydrodynamical) perturbations one gets the shear viscosity $\eta$:
\begin{align}
\eta &= \lim\limits_{\omega\rightarrow 0}\frac{\rho_{(T^\dagger)^{12}T^{12}}(\omega,\ve{k}=0)}{\omega} = \\
&=\intpos{p}\left(\frac{p^1p^2}{\omega}\frac{\partial K(\omega,|\ve{p}|)}{\partial\omega}\rho(\omega,|\ve{p}|)\right)^2\left(-\frac{\partial n(\omega/T)}{\partial\omega}\right). \label{eta0}
\end{align}
This particularly simple expression is a result of the quadratic nature of the EQP-description. There are, however, several examples for calculations done in interacting theories resulting formulae with similar structure \cite{Jeon, transport_YMfrg, transport_NJL3, transport_2PI1, transport_largeN1, transport_largeN2, transport_eff2}. \\
Contrary to Eqs.~(\ref{pressure0}) and (\ref{energy0}), this result cannot be interpreted simply as the sum of viscosities in a mass-distributed gas-mixture. We will see later, that Eq.~(\ref{eta0}) can cover phenomenology beyond the relaxation time approximation. Furthermore, due to the integrable nature of the EQP-action, it is symmetry-preserving, and there is no need of further operator-improvement (e.g.\ by the resummation of vertex corrections as it would be necessary in the 2PI approximation, see for example Ref.~\cite{Jeon}).\\
As a matter of thermodynamic consistency, it turns out, that for a homogeneous and temperature-dependent background the expression Eq.~(\ref{eta0}) is unchanged. For the details of the calculations with non-zero background see Appendix \ref{sourceTerm}.

\section{Non-universal lower bound to $\eta/s$}\label{lowerB}
In the previous sections we have derived quite simple expressions for the entropy density and the shear viscosity in Eqs.~(\ref{pressure0}, \ref{energy0}) and (\ref{eta0}). Using dimensionless quantities, the entropy density over $T^3$ reads as
\begin{align}\label{entropy1}
\sigma :=\frac{s}{T^3} =\intlim{p}{0}{\infty}g(p,T)\chi_s(p/T),
\end{align}
where ${g(p,T)=\frac{\partial K}{\partial p}\overline{\rho}}$, while the thermodynamic weight is
\begin{align}
\chi_s(x) =\chi_{\varepsilon}(x)+\chi_P(x) \approx \frac{x^3}{4\pi^3}K_3(x).
\end{align}
The expression for the shear viscosity contains the very same function $g$:
\begin{align}\label{eta1}
\eta =\intlim{p}{0}{\infty}g^2(p,T)T^4\lambda_\eta(p/T),
\end{align}
with the weight function
\begin{align}
\lambda_\eta(x) =\frac{1}{4\pi^3}\frac{x^5}{15}\intlim{y}{1}{\infty}(-n'(xy))(y^2-1)^{5/2} \approx \frac{x^2}{4\pi^3}K_3(x).
\end{align}
Now we focus on the fluidity measure $\eta/s$, the relaxation coefficient of a transversal hydrodynamical perturbations, as it was mentioned earlier. There is a great interest in theoretical physics whether a universal lower bound to $\eta/s$ exists. It has been theorized in Ref.~\cite{KSSpaper} that this lower bound is $\frac{1}{4\pi}$ in certain conformal field theories with holographic dual. Further investigation showed the possibility of violating this universal value of the lower bound even in the framework of the AdS/CFT duality \cite{KSSviolation1, KSSviolation2} and also in effective theories \cite{nonUnivLowerBnd, nonUnivLowerBnd2, nonUnivLowerBnd3, CohensArgument}. Although we do not expect any universal result in the framework of EQP, the question is still valid. In fact, we are able to give an answer in the EQP-framework. The following variational problem is to be solved:
\begin{align}
\frac{\delta}{\delta g}(\eta[g]-\alpha s[g]) =0,
\end{align}
with the ($p$-independent) Lagrange's multiplier $\alpha$, fixing the value of $s$. Since $s$ is a linear functional of $g$ whilst $\eta$ is quadratic, the solution for the minimizing function is
\begin{align}
g^*(p,T) = \frac{\alpha}{2T}\frac{\chi_s(p/T)}{\lambda_\eta(p/T)}. \label{gstar}
\end{align}
Keeping the value of $s$ fixed, we are able to compute $\eta^*$, the lowest possible value of the shear viscosity in the EQP description depending on the thermodynamic quantities:
\begin{align}
\eta^* =\frac{\alpha^2}{4T^2}\intlim{p}{0}{\infty} \frac{\chi_s^2(p/T)}{\lambda_\eta^2(p/T)}T^4\lambda_\eta(p/T)= \frac{1}{\intlim{y}{0}{\infty}\frac{\chi_s^2(y)}{\lambda_\eta(y)}}\frac{s^2}{T^3}.
\end{align}
Therefore the lower bound to $\eta/s$ with EQP is
\begin{align}
\frac{\eta}{s} \geq \frac{\eta^*}{s}= \frac{1}{\intlim{y}{0}{\infty}\frac{\chi_s^2(y)}{\lambda_\eta(y)}} \sigma =:\frac{\sigma}{I}\approx \frac{1}{48}\sigma.
\end{align}
A speciality of this minimal-$\eta/s$ system is that all the thermodynamic and transport quantities are controlled by $\eta/s=\frac{\sigma}{I}$. The two types of averages we considered in this article are proportional to $\sigma$ or $\sigma^2$, despite a constant tensorial factor:
\[\begin{array}{ccc}
\frac{\avr{T^{\mu\nu}}}{T^4}\sim\sigma, & \mathrm{and} & \frac{\eta}{T^3},\,\frac{\zeta}{T^3},\,\frac{\kappa}{T^3}\sim \sigma^2,
\end{array}\]
$\zeta$ and $\kappa$ being the bulk viscosity and the heat conductivity, respectively.\\

\section{Examples}\label{examples}
Now we turn to analyse counterexamples. The main objective here is to demonstrate the changes in $\eta/s$ while the spectral function interpolates between quasi-particle-like behaviour with narrow peaks and cases with significant continuum contribution.

\subsection{Lorentzian quasi-particle peak}\label{exLorentzian}
\begin{figure}
\centering
  \includegraphics[width=0.9\linewidth]{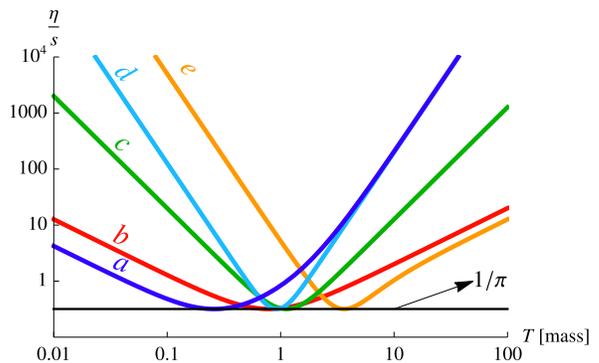}
  \caption{$\eta/s$ versus temperature $T$ provided by the $m=\gamma$ Lorentzian ansatz Eq.~(\ref{BWansatz}). The minimum value is universally $1/\pi$. The plotted lines belong to various choices of $\gamma$ (color online): ${\sim (T^2+T_0^2)^{-1}}$ (blue, $a$), constant (red, $b$),  $\sim T^3$ (green, $c$), $\sim T^{-2}$ (light blue, $d$), ${\sim T^{2+\frac{1}{\epsilon}}}(T_0^\frac{1}{\epsilon}+T^\frac{1}{\epsilon})^{-1}$ with $\epsilon=0.5$ (yellow, $e$).}
\label{fig:lorentzian}
\end{figure}

First we consider a Lorentzian ansatz. It is consistent with the effect of the Dyson resummation in the special case when the self-energy equals to ${~\gamma^2-2\gamma\omega\text{i}}$:
\begin{equation}\label{BWansatz}
\rho_L(\omega,\ve{p})=\frac{4\gamma\omega}{(\omega^2-\ve{p}^2-\gamma^2)^2+4\gamma^2\omega^2}
\end{equation}
The sum-rule ${\frac{1}{2\pi}\intlim{\omega}{-\infty}{\infty}\omega\rho_L(\omega,\ve{p})=1}$ is fulfilled, moreover, ${\rho_L(\omega,\ve{p})\overset{\gamma\rightarrow 0}{\rightarrow} 2\pi\delta(\omega^2-\ve{p}^2)}$. Interestingly, this ansatz is microcausal without any restriction\footnote{Its Fourier-transform is not Lorentz-invariant, but closely related to the free-particle limit $\gamma=0$: $\rho(x)=e^{-\gamma t}\rho_{\gamma=0}(x)$.}. Using Eqs.~(\ref{pressure0}) and (\ref{energy0}) we get:
\begin{equation}
s_L=\frac{1}{4\pi^3}\intlim{\omega}{0}{\infty}2\pi\omega^3\left(-\frac{1}{\omega}\mathrm{ln}(1-e^{-\frac{\omega}{T}})+\frac{1}{T}\frac{1}{e^{\frac{\omega}{T}}-1}\right)= \frac{2\pi^2}{45}T^3,
\end{equation}
where $2\pi\omega^3$ before the parenthesis equals $\int\!\!\mathrm{d}^3\ve{p}\omega\frac{\partial K_L}{\partial\omega}\rho_L$. It coincides, apparently, with the entropy of the ideal Bose gas. For the shear viscosity we evaluate Eq.~(\ref{eta0}):
\begin{equation}
\eta_L= \frac{1}{60\pi^2}\frac{1}{T}\intlim{\omega}{0}{\infty}\left(5\gamma\omega^2+\frac{\omega^4}{\gamma}\right)\frac{1}{\mathrm{ch}\frac{\omega}{T}-1}= \frac{1}{18}\gamma T^2+\frac{2\pi^2}{225}\frac{T^4}{\gamma}.
\end{equation}
Besides the expected $\sim\gamma^{-1}$ term, a linear one appears. The fluidity measure $\eta/s$ reads as:
\begin{equation}\label{etaLor}
\frac{\eta_L}{s_L}= \frac{5}{4\pi^2}\frac{\gamma}{T}+\frac{1}{5}\frac{T}{\gamma}.
\end{equation}
Regardless of the temperature-dependence of $\gamma$, it has the minimal value $\left.\frac{\eta_L}{s_L}\right|_{T^*}=\frac{1}{\pi}$ (Fig.~\ref{fig:lorentzian}). The position of the minimum satisfies the equation ${\gamma(T^*)=\frac{2\pi}{5}T^*}$, for constant $\gamma$ this is $T^*=\frac{5\gamma}{2\pi}$. \\
It is worthwhile to mention, that Eq.~(\ref{etaLor}) is clearly beyond the relaxation time approximation as it has a contribution proportional to the inverse of the quasi-particle lifetime $\sim\gamma$.

\subsubsection*{The long lifetime limit $m\gg\gamma$}\label{exQPapprox}
\begin{figure}
\centering
  \includegraphics[width=0.9\linewidth]{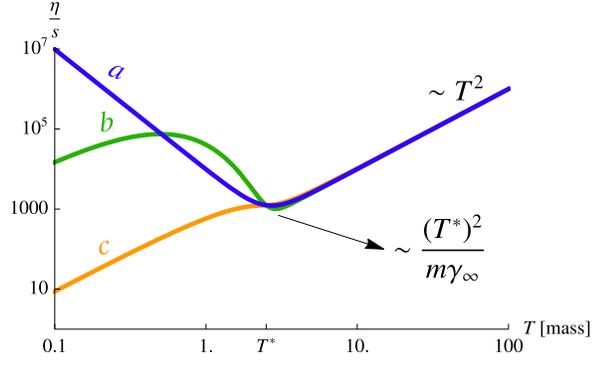}
  \caption{$\eta/s$ in QP-approximation for various $\gamma(T)$ with sudden change at $T^*$. Different low-temperature behaviour of $\eta/s$ are depicted for different $\gamma$-characteristics in the transition region. Exponential relaxation with local minimum and maximum (color online): power-law relaxation with diverging result when $T\rightarrow 0$: ${\sim T^{2-\frac{1}{\tilde{\epsilon}}}((T^*)^{\frac{1}{\tilde{\epsilon}}}+T^{\frac{1}{\tilde{\epsilon}}})}$, $\tilde{\epsilon}=0.2$ (blue, $a$), ${\sim T^2(1+\text{tanh}((T-T^*)/\epsilon))^{-1}}$, $\epsilon=0.5m$ (green, $b$), power-law relaxation with inflexion in $T^*$: ${\sim T^2(1+2/\pi\cdot\text{arctan}((T-T^*)/\epsilon))^{-1}}$ $\epsilon=0.9m$ (yellow, $c$). The value of the minimum is: ${\left.\eta/s\right|_\text{min}=(T^*)^2(m\gamma_\infty)^{-1}+\mathcal{O}(\epsilon)}$, $\gamma_\infty=0.01m$.}
\label{fig:QPapprox}
\end{figure}

In the quasi-particle limit with finite mass $m\gg\gamma$ the Eqs.~(\ref{energy1}), (\ref{pressure1}) and (\ref{eta1}) with the Dirac-delta-approximating spectral function ${\overline{\rho}(p)=2\pi\delta_\gamma(p^2-m^2)}$ result in the following simple expressions:
\begin{eqnarray}
s_{QP} &=& \frac{m^3}{2\pi^2}K_3(m/T), \label{sQP} \\
\eta_{QP} &=& \frac{1}{2\pi^2}\frac{m^2T^2}{\gamma}K_3(m/T). \label{etaQP}
\end{eqnarray}
Here the width of the peak is apparent in $\eta$ only, due to the regularization of the square of the Dirac-delta: ${\delta_\gamma^2(p^2-m^2) \overset{m\gg\gamma}{\approx} \frac{2\pi}{\gamma}\delta(p^2-m^2)}$. The $\eta$ over $s$ ratio reads as
\begin{align}
\frac{\eta_{QP}}{s_{QP}} = \frac{T^2}{\gamma m}.
\end{align}
Let us assume that, for some reason, the particle-lifetime changes significantly around $T=T^*$, but $\gamma\ll m$ still holds (Fig.~\ref{fig:QPapprox}). We parametrize the width as ${\gamma(T) =\gamma_\infty\theta_\epsilon(T-T^*)}$, where $\gamma_\infty$ is its value when $T\gg T^*$ and $\epsilon$ ($\tilde{\epsilon}$) is the size of the transition region in energy dimensions (or in dimensionless units) and $~{m\varepsilon,\,\tilde{\varepsilon}\ll T^*}$ hold. We further assume that $m$ does not change significantly. In case of sharp change in $\gamma$, $\eta/s$ has a well-defined minimum at $T^*+\mathcal{O}(\epsilon)$. Depending on how the transition region is localized, the low-temperature limit of $\eta/s$ could be different. $i)$ When $\gamma$ goes to 0 in an exponential manner, $\eta/s$ reaches zero as $\sim T^2$. $ii)$ If the transition in $\gamma$ is power-law-like: ${\sim (1+(T^*/T)^\frac{1}{\tilde{\epsilon}})^{-1}}$, the ratio is either divergent in $T=0$ or zero:
\begin{equation}
\frac{\eta}{s}\sim \left\{\begin{array}{lll}
T^2, &\mathrm{when} & T\gg T^*, \\
T^{2-\frac{1}{\tilde{\epsilon}}}, &\mathrm{when} & T\ll T^*.
\end{array}\right.
\end{equation}
The minimal value is ${\left.\frac{\eta}{s}\right|_{T^*}\approx\frac{2(T^*)^2}{m\gamma_\infty}}$. \\
A physically realistic situation is when $\gamma\sim T$ and $m \approx\text{const.}$ for $T>m$. In this case the fluidity measure is proportional to $T$ on high temperature.

\subsection{Quasi-particle and its continuum tail}\label{exQPtail}
\begin{figure}
\centering
  \includegraphics[width=0.9\linewidth]{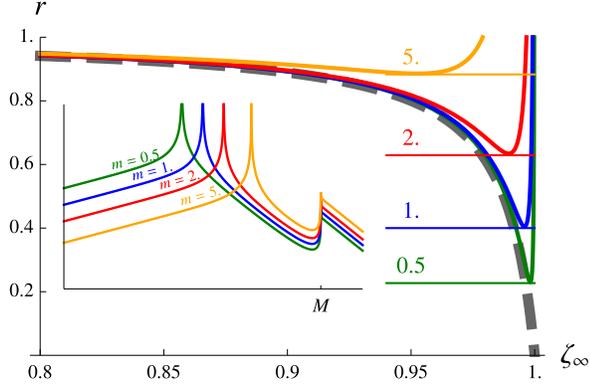}
  \caption{The ratio $r=\frac{\eta/s}{\eta_{QP}/s_{QP}}$ versus $\zeta_\infty$. Fixing $\frac{\gamma_p}{m}=0.05$, $M=50.0m$, $\epsilon=0.015m$ and $T=100m$, graphs with various values of $m$ (0.5, 1.0, 2.0, 5.0) are plotted, so that ${m,\,M\gg\epsilon}$ holds. For given $m$, $M$ and $\gamma_p$ the sum rule provides: ${\Gamma\approx \frac{\pi}{4}\left(\frac{1}{2}-\frac{1}{\pi}\arctan\frac{m}{\gamma}\right)\frac{M}{\zeta_\infty}}$ for $m\ll M$. The dashed line indicates the limiting case $M\rightarrow\infty$. An illustration of the corresponding spectral functions are inset on a double-logarithmic plot, at $\zeta=0.9$}
\label{fig:QPtail}
\end{figure}

We move towards to more general situations and parametrize the retarded propagator with momentum dependent self-energy: $~{m^2(p)-p\gamma(p)\text{i}}$ and wave-function renormalization $Z(p)$
:
\begin{equation}
G^{ra}(p)=\frac{Z(p)}{p^2-m^2(p)+ip\gamma(p)}.
\end{equation}
First we assume $\gamma(p)$ and $Z(p)$ to be analytic functions and keep $m$ constant. The kernel function then reads as follows:
\begin{align}
g(p) &= \frac{\partial K}{\partial p}\overline{\rho}(p) =\frac{\left(2p-(p^2-m^2)\frac{Z'(p)}{Z(p)}\right)p\gamma(p)}{(p^2-m^2)^2+p^2\gamma^2(p)} =\\
&=: g_\text{peak}(p)-\frac{p\gamma(p)(p^2-m^2)\frac{Z'(p)}{Z(p)}}{(p^2-m^2)^2+p^2\gamma^2(p)}=g_\text{peak}(p)+g_\text{cont}(p), \nonumber
\end{align}
where we separated the Lorentzian \textit{peak contribution}. The remaining \textit{continuum} part bears the same pole structure as $g_\text{peak}$ but with $~{p^2-m^2}$ in the nominator also, and therefore disappears in the $\gamma\rightarrow 0$ limit. Using $Z(p)$ to cut out the $p<M$ part of the continuum $g_\text{cont}$ with $m<M$, we are left with an $\mathcal{O}(\gamma)$ contribution for constant $\gamma$. \\
Keeping in mind that we interested in going beyond the QP-spectrum in a parametrically controlled way, we link $Z$ and $\gamma$ together. For $Z=1$ and $\gamma=0$ we expect the particle excitation to be restored with mass $m$ and with infinite lifetime. We force $Z<1$ and $\gamma>0$ to happen simultaneously by setting ${\gamma\overset{!}{=}\Gamma(1-Z(p))=:\Gamma\zeta(p)}$ with a constant $\Gamma$ with dimension of energy. To ensure that $Z<1$ is restricted to $~{p>M>m}$, we put ${\zeta(p)=\zeta_\infty\theta_\epsilon(p-M)}$, where $0<\zeta_\infty<1$ and $\epsilon$ encodes how sudden the change from 0 to $\zeta_\infty$ is. If $~{M\gg\epsilon}$, the integrals in Eqs.~(\ref{entropy1}) and (\ref{eta1}) pick the $p\approx M$ contributions only, resulting in
\begin{align}
s \approx & s_{QP}(m,T) + \frac{\frac{\zeta_\infty}{1-\zeta_\infty}\Gamma M(M^2-m^2)}{(M^2-m^2)^2+\zeta_\infty^2\Gamma^2M^2}s_{QP}(M,T), \\
\eta \approx &\eta_{QP}(m,T,\gamma_p) + \nonumber \\
  &+ \frac{\frac{\zeta^2_\infty}{1-\zeta_\infty}\Gamma^2M^2(M^2-m^2)\left(4\epsilon M+\frac{M^2-m^2}{1-\zeta_\infty}\right)}{\left[(M^2-m^2)^2+\zeta_\infty^2\Gamma^2M^2\right]^2}\eta_{QP}(M,T,\epsilon),
\end{align}
with $s_{QP}$, $\eta_{QP}$ defined by Eqs.~(\ref{sQP}) and (\ref{etaQP}), respectively. ${\gamma_p=\Gamma\zeta(m)\ll\Gamma}$ and $\epsilon$ are present to regularize the $\delta^2$-like parts in the viscosity integral. Writing out $\eta$ over $s$ explicitly:
\begin{align}
\frac{\eta}{s} &\approx \frac{\eta_{QP}(m,T,\gamma_p)}{s_{QP}(m,T)}\frac{1+A^2(m,M,\Gamma,\zeta_\infty)\frac{\eta_{QP}(M,T,\epsilon)}{\eta_{QP}(m,T,\gamma)}}{1+A(m,M,\Gamma,\zeta_\infty)\frac{s_{QP}(M,T)}{s_{QP}(m,T)}} =\nonumber \\
&= \frac{T^2}{m\gamma_p}\frac{1+A^2(m,M,\Gamma,\zeta_\infty)\frac{M^2}{m^2}\frac{\gamma_p}{\epsilon}\frac{K_3(M/T)}{K_3(m/T)}}{1+A(m,M,\Gamma,\zeta_\infty)\frac{M^3}{m^3}\frac{K_3(M/T)}{K_3(m/T)}}, \,\, \text{with}\\
&A(m,M,\Gamma,\zeta_\infty) = \frac{\frac{\zeta_\infty}{1-\zeta_\infty}\Gamma M(M^2-m^2)}{(M^2-m^2)^2+\zeta_\infty^2\Gamma^2M^2} \nonumber. 
\end{align} 
The above expression results in a reduced value of ${\eta/s}$ compared to ${\eta_{QP}(m,T,\gamma_p)/s_{QP}(m,T)}$ whenever ${\zeta_\infty<(1+\frac{1}{2}\frac{m}{M}\frac{\Gamma}{\epsilon}\theta_\epsilon(m-M))^{-1}}$ holds. 
The ratio ${r=\frac{\eta/s}{\eta_{QP}/s_{QP}}}$ has a minimal value ${2\frac{m}{M}\frac{\gamma_p}{\epsilon}\left(\sqrt{1+\frac{M}{m}\frac{\epsilon}{\gamma_p}}-1\right)}$ for $T\gg M$. \\
Naively, one would think that the continuum contributions are suppressed for large $M$. Nevertheless $Z(p)$ and $\gamma(p)$ are momentum-dependent, and the sum rule $~{\frac{1}{\pi}\intlim{p}{0}{\infty}p\overline{\rho}(p)=1}$ imposes a constraint on the parameters. $\Gamma$ happens to be proportional to $\frac{M}{\zeta_\infty}$, therefore $r$ is not the trivial $~{r=1}$ in the large-$M$ limit. Near $\zeta_\infty\approx 1$ its value drops considerably, see Fig.~\ref{fig:QPtail} for examples. 
Consequently, the fluidity measure is modified by the "continuum" parameters $M$ and $\epsilon$ and reaches its minimal value when $\zeta_\infty\approx 1$:
\begin{align}
\left.\frac{\eta}{s}\right|_\text{min.} &\overset{\frac{M}{T}\ll 1}{\approx} \frac{2T^2}{M\epsilon}\left(\sqrt{\frac{M\epsilon}{m\gamma_p}+1}-1\right) =\nonumber \\
&\overset{\epsilon =\gamma_p}{=} \frac{T^2}{\gamma_p m}\underbrace{\frac{2m}{M}\left(\sqrt{\frac{M}{m}+1}-1\right)}_{\leq 1} \overset{\frac{m}{M}\ll 1}{\approx} \frac{2T^2}{\gamma_p \sqrt{mM}}
\end{align}

\begin{figure}
\centering
	\includegraphics[width=0.7\linewidth]{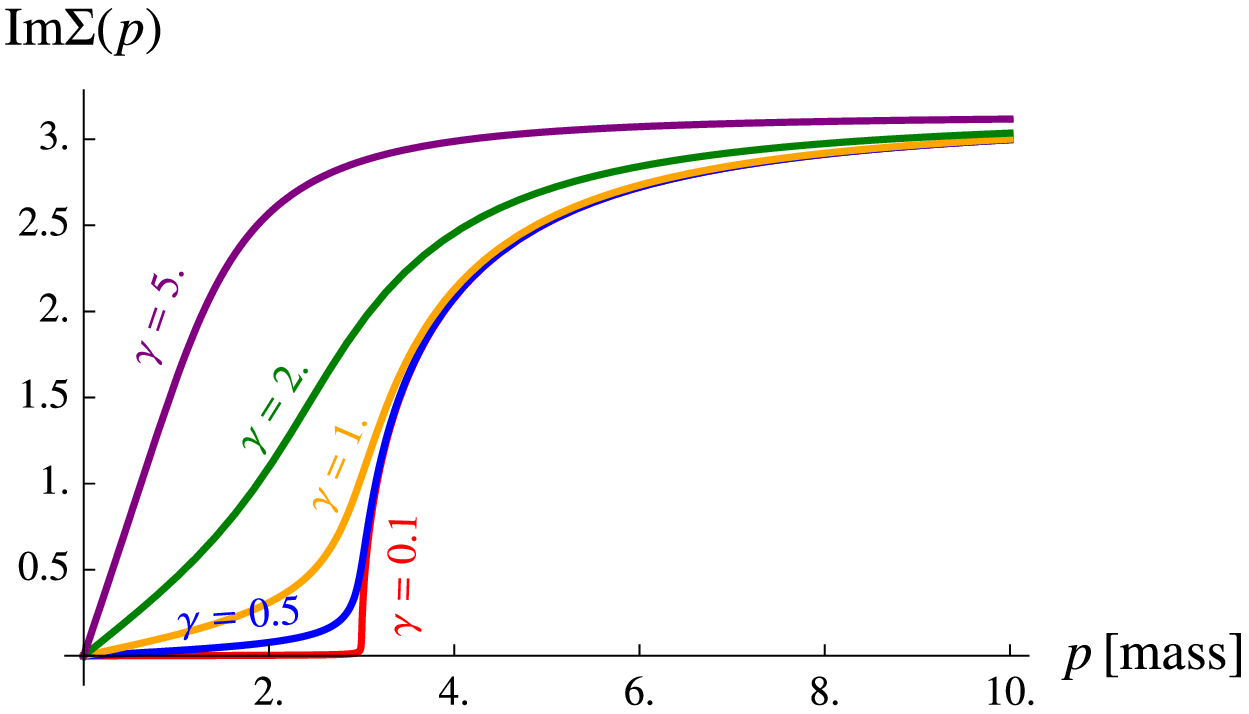}\,
	\includegraphics[width=0.7\linewidth]{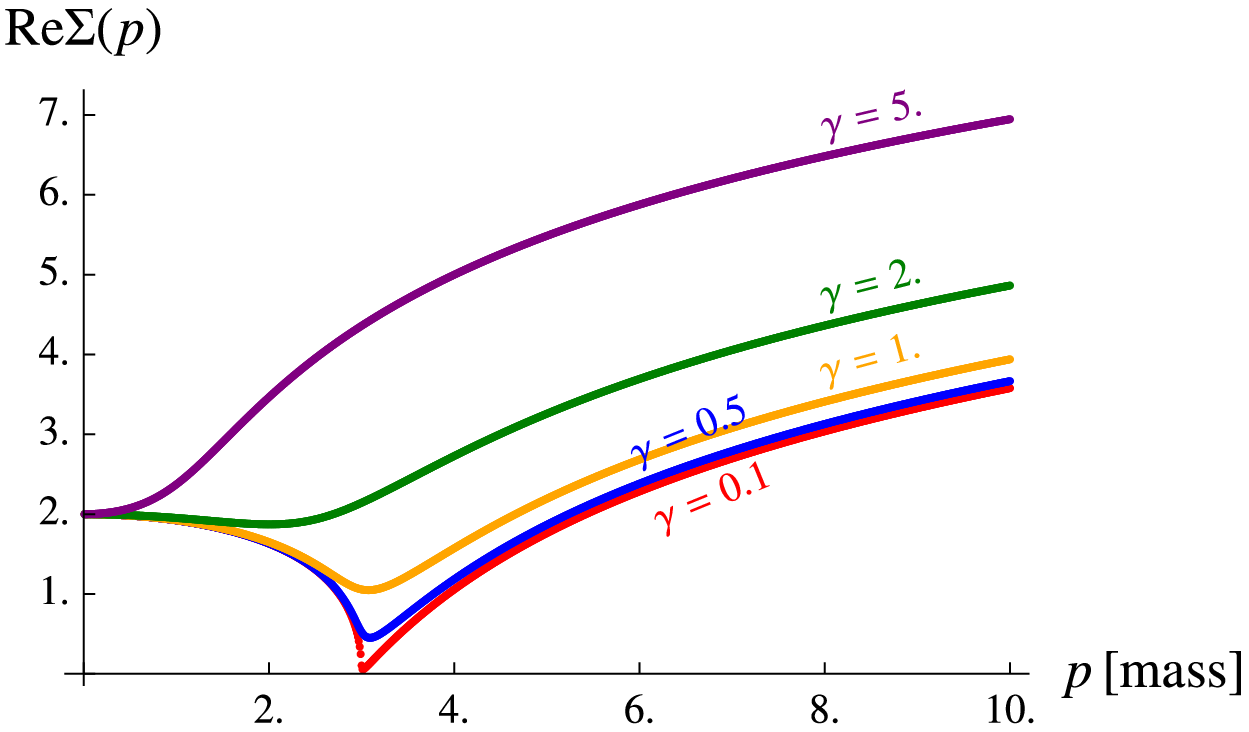}
\caption{Imaginary and real parts of the self energy $\Sigma_s$ on the real line, with parameters $M=3.0$ (in the dimension of mass) and $\zeta=1.0$ (in the dimension of mass square).}
\label{fig:Sigma}
\end{figure}
\begin{figure}
\centering
  \includegraphics[width=0.7\linewidth]{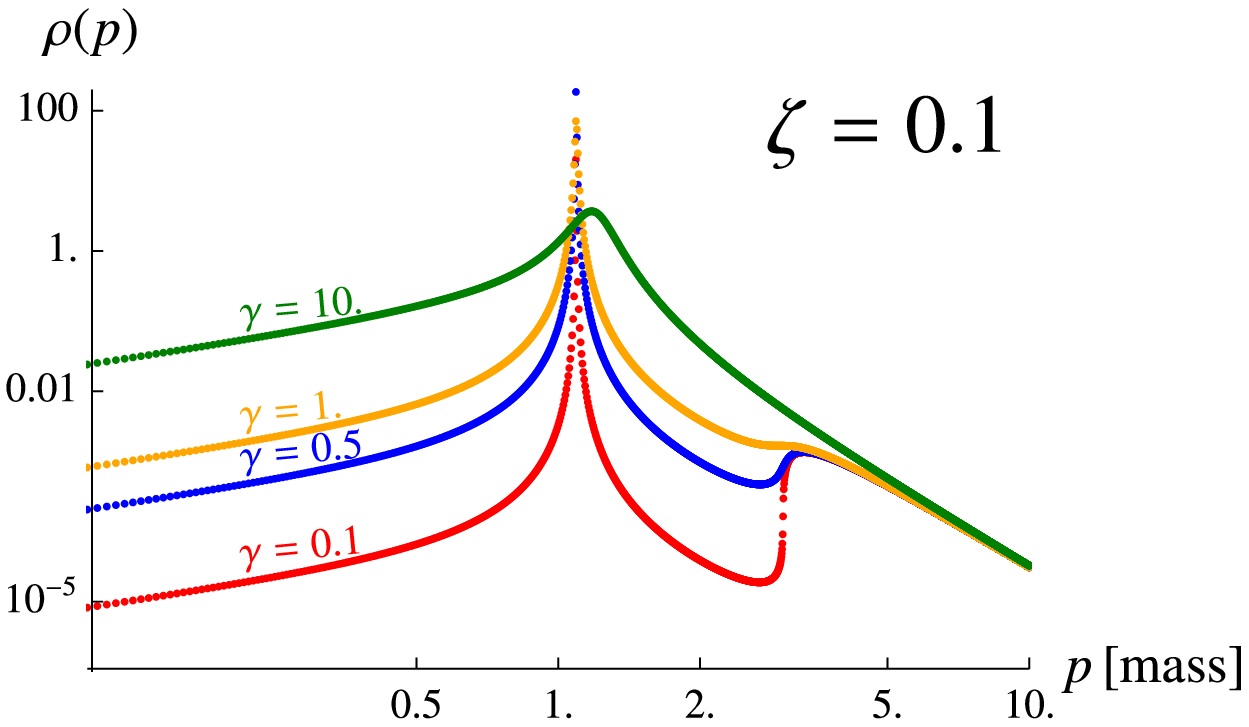}
  \includegraphics[width=0.7\linewidth]{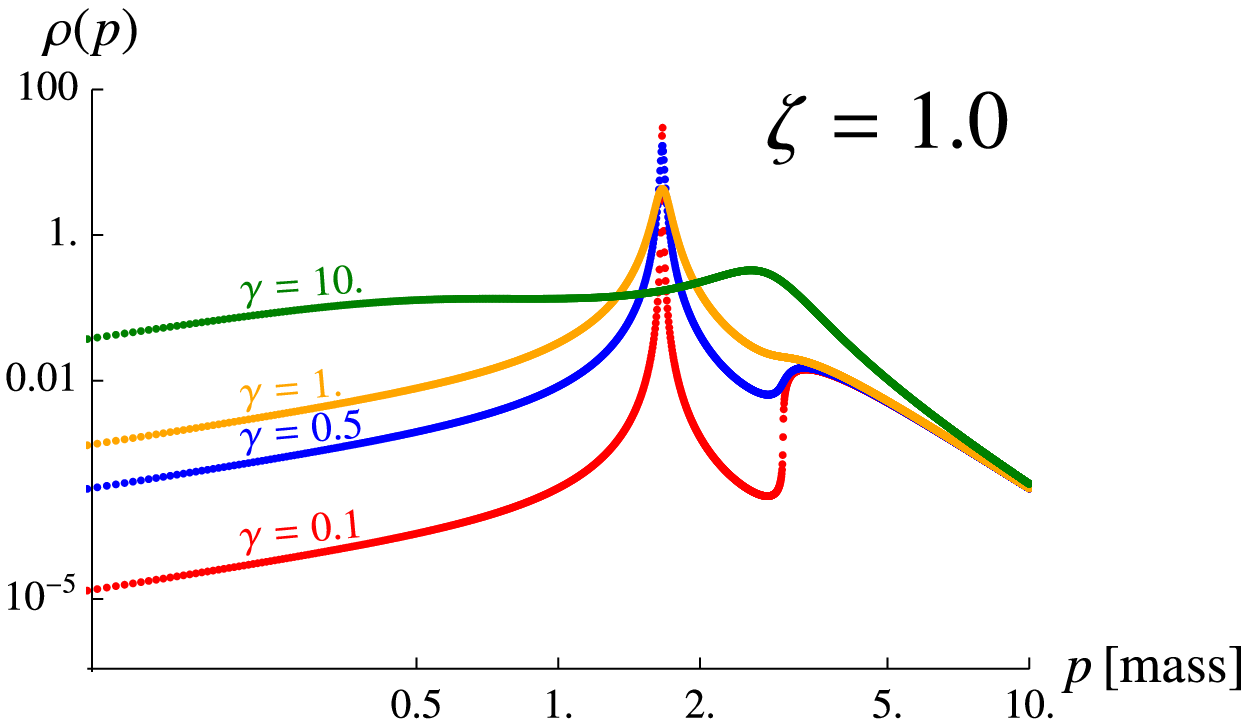}
  \includegraphics[width=0.7\linewidth]{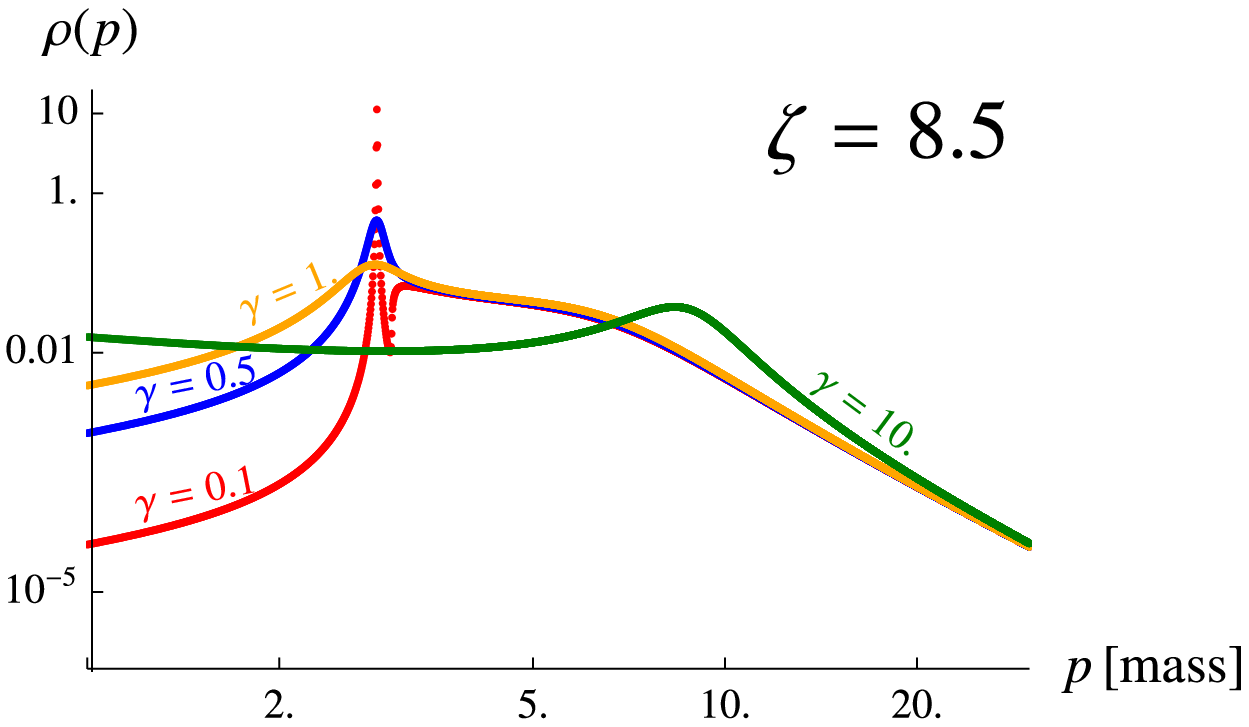}
  \caption{Spectral density $\overline{\rho}(p)$ for various values of ${~\gamma=0.1-10.0}$, with fixed parameters $m=1.0$, $M=3.0$ (in the dimension of mass) and ${~\zeta=0.1,\,1.0,\,8.5}$ (in the dimension of mass square). After the pole-part and the continuum ''melted'' into each other ($\gamma\approx 1.0$), the further increase of $\gamma$ shifts the QP-peak towards higher momenta.}
\label{fig:cutSigmaRho}
\end{figure}
\begin{figure}
\centering
  \includegraphics[width=0.9\linewidth]{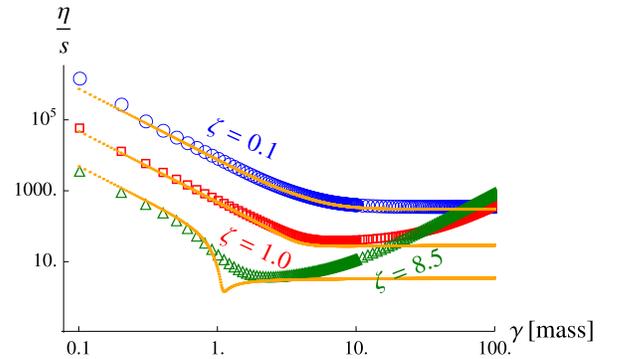}
  \caption{The fluidity measure $\eta/s$ computed with a realistic ansatz for the self energy $\Sigma_s$. The ratio $\eta/s$ reduces as the continuum contribution is more and more pronounced by the increase of $\zeta$. The $\gamma$-dependence shows a minimal value of $\eta/s$, far from the region where the QP-peak and the continuum part are well distinguishable. The QP-pole approximation of Eq.~(\ref{QPpoleAppr}) is indicated by the thin curves. The values of other fixed parameters are $m=1.0$, $M=3.0$ and $T=10.0$.}
\label{fig:etaOsSigma}
\end{figure}

\subsection{Beyond the QP-pole}\label{exCutSigma}
As we have seen, if $G^{ra}$ has only pole singularities, those control the overall behaviour of the theory inevitably. Mimicking the features of the multi-particle contribution using the QP-tail is inadequate in the sense that its effect is suppressed by the imaginary part of the pole position: the width of the QP-peak. In more realistic situations, i.e.\ in interacting QFTs, the propagator has branch cuts beside its poles. Branch cuts are generated even in one-loop order in perturbation theory, corresponding to the opening of multi-particle scattering channels. For example, in a theory with the lowest mass excitation $m$, the continuum contribution of the spectrum starts at $M=2m$ (at zero temperature, if {~1-to-2} decay or {~2-to-2} scattering is allowed at tree-level). To take into account these cut contributions we parametrize the inverse retarded propagator and the spectral function as follows: 
\begin{align}
(G^{ra})^{-1} &= p^2-m^2-\Sigma_s, \label{invGCut} \\
\overline{\rho} &= \frac{\text{Im}\Sigma_s}{(p^2-m^2-\text{Re}\Sigma_s)^2+(\text{Im}\Sigma_s)^2}. \label{rhoCut}
\end{align}
At zero temperature, we assume $\Sigma_s$ to have a branch cut along the real line, starting at $p=M$. At finite temperature we expect the near-$M$ behaviour of $\im\Sigma_s$ smoothens. We use an ansatz that shows this kind of behaviour. It is motivated by the self-energy correction of a cubic scalar model (see for eample Sec.~24.1.1 of Ref.~\cite{SchwartzQFT}) and by the IR-safe resummation discussed in Ref.~\cite{GenBoltzmannEqu}:
\begin{align}
\im\Sigma_s(p) &=\zeta \pi\frac{\sqrt{\sqrt{\left(1-\frac{M^2}{p^2}\right)^2+4\frac{\gamma^4}{M^4}}+1-\frac{M^2}{p^2}}}{\sqrt{\sqrt{1+4\frac{\gamma^4}{M^4}}+1}} \\
&\overset{\gamma\rightarrow 0}{\longrightarrow} \zeta\pi\theta(p-M)\sqrt{1-\frac{M^2}{p^2}},\nonumber\label{cutImSigma}
\end{align}
\begin{align}
\re\Sigma_s(p) &=\frac{1}{\pi}\mathcal{P}\!\!\!\!\!\!\! \intlim{q}{0}{\infty}\frac{2q\im\Sigma_s(q)}{p^2-q^2} \\
&\overset{\gamma\rightarrow 0}{\longrightarrow} \left\{ \begin{array}{lll}
	2\zeta\sqrt{\frac{M^2}{p^2}-1}\cdot\arcsin\left(\frac{p}{M}\right), & \,\, & p<M,\\
	-2\zeta\sqrt{1-\frac{M^2}{p^2}}\cdot\text{ln}\left(\frac{p}{M}-\sqrt{\frac{p^2}{M^2}-1}\right), &\,\, & M<p.
\end{array}\right. \label{cutReSigma}
\end{align}
The Kramers-Kronig relation is used to evaluate $\re\Sigma_s$ for any values of $\gamma$ numerically. We plotted the self-energy $\Sigma_s$ and the spectral density $\overline{\rho}$ on Figs. \ref{fig:Sigma} and \ref{fig:cutSigmaRho} for illustration. The limit $\gamma=0$ is also given analytically in Eq.~(\ref{cutReSigma}). \\
Formulae in Eqs.~(\ref{entropy1})~and~(\ref{eta1}) are used to evaluate the fluidity measure $\eta/s$. The numerical results are depicted on Fig.~\ref{fig:etaOsSigma} for various values of $\gamma$ with fixed $\zeta$. The main conclusion here is that the increase of the weight of the continuum in $\overline{\rho}$ by increasing the value of $\zeta$, the ratio $\eta/s$ decreases. As for the $\gamma$-dependence of the fluidity measure, we find a power-law-like decay ending in a minimum. This decrease of $\eta/s$ seems to be connected to the ''melting'' of the QP-peak and the multiparticle continuum in $\overline{\rho}$. Leaving this region of the parameter space, i.e.\ further enhancing $\gamma$, the ratio saturates, than starts to slowly increase, see Fig.~\ref{fig:etaOsSigma}. This is mainly the result of the shifting of the QP-peak towards to higher momenta, cf.\ Fig.~\ref{fig:cutSigmaRho}. For comparison, we also show solely the contribution of the QP-peak on Fig.~\ref{fig:etaOsSigma} (indicated by the thin curves), which is calculated by using the formula:
\begin{align}
\left.\frac{\eta}{s}\right|_{\text{QP-pole}} &= \frac{T^2}{2M_*}\frac{2M_*-\frac{\partial\re\Sigma_s(M_*)}{\partial p}}{\im\Sigma_s(M_*)}, \label{QPpoleAppr}
\end{align}
where the pole-mass $M_*$ satisfies the equation: $~{M_*^2-m^2-\re\Sigma_s(M_*)=0}$. This approximation of $\eta$ and $s$ becomes worse and worse with increasing the value of $\zeta$, as it is expected.

\subsection{On phase transition} \label{nearCEP}
Hitherto we investigated systems whose thermodynamical quantities were continuous functions of the temperature. We argued that our framework may tackle the phenomenology in the crossover-region, near to a possible critical end point (CEP), where the long-range correlations play an important role. Let us now make here few remarks on the issue of phase transition. \\
As we mentioned earlier, it was observed in a wide range of materials with a CEP in their phase diagrams, that $\eta/s$ shows a considerable reduction near the critical temperature $T_c$. We note here two jointly present effects, both which can contribute to the behaviour of $\eta/s$ as a function of the temperature near to $T_c$. The dimensionless entropy density $s/T^3$ changes more and more sharply approaching the critical temperature. It saturates to the Stefan--Boltzmann-limit for high $T$ and vanishes by lowering the temperature. Therefore, depending on the details of the transition, $T^3/s$ could show significantly different behaviour below and above $T_c$ -- even possibly diverge for $T\rightarrow 0$. That in itself is enough to develop a minimum for $\eta/s$ , even if $\eta/T^3$ is monotonous. Crossing a a 1$^\text{st}$ order type phase boundary, the value of $T^3/s$ jumps, whilst for a 2$^\text{nd}$ order transition its slope is refracted. \\
Besides, $\eta/T^3$ may also tend differently as a function of temperature above and below a characteristic value of $T^*$. We refer to Eq.~(\ref{etaLor}) as a simple example. Although it depends smoothly on temperature for constant $\gamma$, a jump or refraction of the slope is conceivable whenever the temperature dependence of $\gamma$ changes passing the critical temperature. The value of $T^*$ characterizing this transition point is expected to be close to the critical temperature of the system, $~{T_c/T^*\approx\mathcal{O}(1)}$. In case of the Lorentzian for constant $\gamma$ this temperature value is in the order of $\gamma$, namely $~{T^*=\frac{5\gamma}{2\pi}}$. \\
In fluids, it is observed that $\eta$ acts like a susceptibility and diverges weakly as the correlation length $\xi$ goes to infinity. The critical exponent of the shear viscosity is reported to be very small compared to those of the correlation length \cite{etaOsCEP1, etaOsCEP2, etaOsCEP3}. We can use Eq.~(\ref{etaLor}) again, with the tentative identification $~{\gamma\sim\xi^{-1}}$, where $\xi$ is the correlation length (since $\gamma$ is also the mass parameter in the example of Sec.~\ref{exLorentzian}). This would result in a critical behaviour $~{\eta/s\sim |T-T_c|^{-\nu}}$, i.e. the critical exponents of $\xi$ and $\eta/s$ would be the same. This value of the critical exponent is way to high compared to the experimental findings. It is worth to emphasize though, that our approach is based on the Gaussian approximation of the generating functional. Therefore it is not expected to describe the phenomenology in the CEP, where the fluctuations of the order parameter are huge.

\section{Conclusions and outlook}\label{dissCon}
In this paper, we investigated how the robust properties of the spectral density of states $\rho$ of a QFT define the value of the fluidity measure $\eta/s$ in the framework of extended quasi-particles. Without other conserved charges, this ratio characterizes the relaxation to thermal equilibrium after a small shear stress is applied. We worked out formulae both for thermal quantities and transport coefficients in the linear response regime regarding an approximation scheme parametrized solely by $\rho$. This scheme is able to incorporate finite lifetime effects and multi-particle correlations caused by interaction. \\
Parametrizing $\rho(p)$ by microscopically meaningful quantities like the inverse lifetime and mass of quasi-particle excitations (position of the pole singularity of $\rho(p)$), multi-particle threshold (position of the branch point of $\rho(p)$) we analysed the fluidity measure $\eta/s$. Our main finding is that the more non-quasi-particle-like $\rho$ is, the more fluent the medium it describes. More precisely, we tuned the parameters of the spectral function $\rho$ in such a way that the strength or residuum of the quasi-particle peak became less and less pronounced, and we observed the reduction of $\eta/s$. All-in-all, the particularly simple formula of Eq.~(\ref{etaLor}) has proven to be very insightful, especially in the light of the more complicated examples, since it seems to be showing all the key features we have explored during the analysis done in Sec.\ \ref{exQPtail}~and~\ref{exCutSigma}. \\
Our result supports the observations of other authors. The weakening of $\eta$ is also observed in resummed perturbation theory of the quartic interacting scalar model \cite{Jeon}, and also supported by numerical evidences in case of hadronic matter when one takes into account a continuum of Hagedorn-states besides the hadronic resonances \cite{transport_greiner}. \\
We pointed out, that in our framework there is a lower bound to $\eta/s$, which is proportional to the entropy density over $T^3$. As long as one can constrain the thermodynamic quantities, our approach provides a restriction to the transport. \\
Moreover, the approximation of the transport coefficients is feasible based on the detailed knowledge about the thermal observables. Supposing that one knows all the independent thermodynamic quantities as a function of some control parameter (e.g.\ temperature), there is room for a model with as many parameters as the number of the independent thermal observables. Fitting the formulae to the known data set, the parameters $\alpha_i(T)$ in ${g(p,\{\alpha_i(T)\})=\frac{\partial K}{\partial p}\overline{\rho}}$ can be fixed. Therefore the viscosity in the framework of EQP is determined, using $g^2(p)$ and the formula (\ref{eta1}). There is available data from lattice Monte-Carlo simulations describing observables in thermal equilibrium in QCD, and also from condensed matter systems and other field theories. However, it is still challenging to extract the transport coefficients. The estimation based on thermal observables can be a good guideline here.\\
Since more independent thermodynamical quantities mean more conserved charges (besides the energy-momentum density), the formulae given here need to be generalized. The first straightforward step into this direction is to consider the cases of the charged scalar field and the Dirac-field. It would be also interesting to see how the lower bound on $\eta/s$ changes when the chemical potential corresponding to the charge density comes into play. These subjects, however, are left to be discussed in future publications.

\begin{acknowledgements}
Discussions with T. S. Bir\'o, P. V\'an and P. Mati are gratefully acknowledged. We thank M. Vargyas, P. Mati and T. S. Bir\'o for critically reading the manuscript. This work has been supported by the Hungarian National Science Fund OTKA under the contract numbers K104260, K104292 and K116197. MH was also supported by the ''Young Talents of the Nation Scholarship'' \mbox{NTP-EF\"O-P-15} of the Human Capacities Grant Management Office, Hungary. 
\end{acknowledgements}

\appendix
\subsection*{Notations}
Throughout this appendices, the lower index for a space-time or momentum-space dependent quantity means its argument, i.e.\ a four-vector: 
\begin{align}
\varphi_x\equiv\varphi(x)\equiv\varphi(x^0,\ve{x}).\nonumber
\end{align}
Also an integral sign with lower indexed variable of integration is prescribed on the whole domain of the variable (space or momentum-space):
\begin{align}
\int_p(\dots) &= \frac{1}{(2\pi)^4}\intlim{p^0}{-\infty}{\infty}\intlim{^3\ve{p}}{}{}(\dots),\,\, \text{in momentum space,} \nonumber \\
\int_x(\dots) &= \intlim{x^0}{-\infty}{\infty}\intlim{^3\ve{x}}{}{}(\dots),\,\, \text{in space.} \nonumber
\end{align}

\section{Propagators}\label{propagators}
We briefly summarize here the relations between propagators and expectation values that will be useful later on. The numerical upper indices are Keldysh-indices of the real-time formalism. The Keldysh-propagators are defined as $i\mathcal{G}_{x,y}^{ab}=\avr{\mathcal{T_C}\kelphid{x}{a}\kelphi{y}{b}} \equiv \avr{\kelphid{x}{a}\kelphi{y}{b}}$. From now on we omit $\mathcal{T_C}$ which represents the time-ordering on the Keldysh-contour $\mathcal{C}$ \cite{keldysh_ctp}. We also suppress the adjoint sign since in the case of real scalar fields it is equivalent with the identification $\varphi^\dagger_p=\varphi_{-p}$. \\
In general, the following identities hold between the propagators:
\begin{align}
i\mathcal{G}_{x,y}^{11} &=\theta(x^0-y^0)i\mathcal{G}^{21}_{x,y}+\theta(y^0-x^0)i\mathcal{G}^{12}_{x,y}, \\ 
i\mathcal{G}_{x,y}^{22} &=\theta(x^0-y^0)i\mathcal{G}^{12}_{x,y}+\theta(y^0-x^0)i\mathcal{G}^{21}_{x,y}, \\
0 &=\mathcal{G}^{12}_{x,y}+\mathcal{G}^{21}_{x,y}-\mathcal{G}^{11}_{x,y}-\mathcal{G}^{22}_{x,y},\\
\rho_{x,y} &=i\mathcal{G}^{21}_{x,y}-i\mathcal{G}^{12}_{x,y}.
\end{align}
In thermal equilibrium, the propagators are translational invariant, thus for the Fourier-transformed ones ${iG_{p,q}^{ab} \equiv \delta_{p-q}iG_p^{ab}}$. We denote the Fourier-transform of a space-time dependent quantity calligraphic $\mathcal{G}$ with an italic $G$. Furthermore,
\[\begin{array}{ccccc}
iG^{12}_p=n_p\rho_p, & & iG^{21}_p=(n_p+1)\rho_p, & & n_p=\frac{1}{e^{\omega/T}-1}.
\end{array}\]
\begin{widetext}
The following parity-relations hold:
\[\begin{array}{ccccccc}
iG^{12}_{-p}=iG^{21}_p, & & iG^{11}_{-p}=iG^{11}_p, & & iG^{22}_{-p}=iG^{22}_p, & & \rho_{-p} =-\rho_p.
\end{array}\]
In the special case of quadratic action, Wick's theorem holds with the Keldysh-indices signed properly. We need the four-point function for the viscosity calculation:
\begin{align}
\avr{\kelphi{p}{a}\kelphi{q}{b}\kelphi{r}{c}\kelphi{s}{d}} =& \avr{\kelphi{p}{a}\kelphi{q}{b}}\avr{\kelphi{r}{c}\kelphi{s}{d}} + \avr{\kelphi{p}{a}\kelphi{r}{c}}\avr{\kelphi{q}{b}\kelphi{s}{d}}+ \avr{\kelphi{p}{a}\kelphi{s}{d}}\avr{\kelphi{q}{b}\kelphi{r}{c}}. \label{fourpiontfunc}
\end{align}

\section{Energy-momentum tensor}\label{enmomTensor}
We discuss the detailed derivation of the energy-momentum tensor in case of a non-local quadratic action. First we translate $\varphi$ by a space-time dependent field $\alpha$. The variation of the action respect to $\alpha$ provides us the gradient of the energy-momentum tensor (when $\alpha\rightarrow 0$): 
\begin{align}
\left.\Int{x}\frac{\delta S[e^{\alpha\partial}\varphi]}{\delta\alpha^\mu_x}\alpha_x\right|_{\alpha\equiv 0} &= -\Int{x}\alpha_x(\partial_x\cdot T_x)^\mu =\frac{\mathrm{d}}{\mathrm{d}\varepsilon}\left[\frac{1}{2}\Int{x}\varphi_{x^\mu+\varepsilon\alpha_x^\mu}\Int{z}\mathcal{K}_ze^{z\cdot\partial_x}\varphi_{x^\mu+\varepsilon\alpha_x^\mu}
\right]_{\varepsilon=0} =\\
&= \frac{1}{2}\Int{x}\alpha_x\partial_x^\mu\varphi_x\Int{z}\mathcal{K}_ze^{z\cdot\partial_x}\varphi_x +\frac{1}{2}\Int{x}\varphi_x\Int{z}\mathcal{K}_ze^{z\cdot\partial_x} \alpha_x\partial^\mu_x\varphi_x.
\end{align}
After Fourier-transform $\varphi$, $\varphi^\dagger$, $T^{\mu\nu}$ and $\alpha$ one gets
\begin{align}
\Int{k}\alpha_k(-ik_\nu T^{\mu\nu}_{-k}) =& \frac{1}{2}\Int{k}\alpha_k\Int{p}\Int{q}\varphi_p^\dagger\varphi_q \Int{x}\Int{z}\left[ -e^{ik\cdot z}ip^\mu e^{-ip\cdot x}\mathcal{K}_ze^{z\cdot\partial_x}e^{iq\cdot x} +e^{-ip\cdot x}\mathcal{K}_ze^{z\cdot\partial_x}e^{ik\cdot x}iq^\mu e^{i\cdot x}\right],
\end{align}
where the difference between the field variable and its conjugate is indicated. Using the identity $\delta_{k+p-q}=\delta_{k+p-q}\frac{k\cdot(p+q)}{q^2-p^2}$ and collecting the terms result in
\begin{align}
ik_\nu T^{\mu\nu}_k =& \frac{1}{2}\Int{p}\Int{q}\varphi_p^\dagger\varphi_q \delta_{k+p-q}\left(ip^\mu K_q-iq^\mu K_p \right)= \frac{1}{2}\Int{p}\Int{q}\varphi_p^\dagger\varphi_q \delta_{k+p-q}\left( ip^\mu\frac{k\cdot(p+q)}{q^2-p^2}( K_q- K_p) -ik^\mu K_p\right) = \label{preDkernel} \\
=& \frac{1}{2}\Int{p}\Int{q}\varphi_p^\dagger\varphi_q \delta_{k+p-q} ip^\mu\frac{k\cdot(p+q)}{q^2-p^2}( K_q- K_p) =: ik_\nu \frac{1}{2}\Int{p}\Int{q} \varphi_p^\dagger\varphi_q \delta_{k+p-q}D^{\mu\nu}_{p,q}. \label{Dkernel} 
\end{align}
Here in Eq.~(\ref{Dkernel}) we left the last term in the parenthesis of Eq.~(\ref{preDkernel}). This can be done because of the EoM ${~K_p\varphi_p=0}$. Averaging the non-$k$-orthogonal part of $T_k^{\mu\nu}$ over the equilibrium ensemble, we arrive the energy-momentum density $\varepsilon^{\mu\nu}$ (in what follows, we subtract the divergent terms proportional to the volume of the system):
\begin{align}
\varepsilon^{\mu\nu} =& \Int{k}\avr{T_k^{\mu\nu}}=\frac{1}{2}\Int{k}\Int{p}\Int{q}\avr{(\varphi_p^\dagger)^{(1)}\varphi^{(2)}_q} \delta_{k+p-q}D^{\mu\nu}_{p,q}= \frac{1}{2}\Int{k}\Int{p}\Int{q}iG^{12}_p\delta_{p-q} \delta_{k+p-q}D^{\mu\nu}_{p,q}=\frac{1}{2}\Int{p}D^{\mu\nu}_{p,p}\rho_p n_p, \\
D^{\mu\nu}_{p,p} =& \lim\limits_{q\rightarrow p} \frac{p^\mu(p+q)^\nu}{q^2-p^2}(K_q- K_p) \overset{q=p+\zeta n}{=} \frac{p^\mu p^\nu}{n\cdot p}\lim\limits_{\zeta\rightarrow 0}\frac{ K_{p+\zeta n}- K_p}{\zeta} \overset{K_p\equiv K_{|p|}}{=} \frac{p^\mu p^\nu}{|p|}\frac{\partial K_{|p|}}{\partial |p|} =\frac{p^\mu p^\nu}{\omega}\frac{\partial K_{|p|}}{\partial\omega}. 
\end{align}

\section{Shear viscosity}\label{visco}
Using the definition of the spectral function of an operator, we derive $\rho_{T^\dagger T}$. With the renormalized energy-momentum tensor in Eq.~(\ref{Dkernel}) and using the relation in Eq.~(\ref{fourpiontfunc}) the computation is straightforward:
\begin{align}
\rho_{(T^\dagger)^{ij}T^{ij},k} =& iG^{21}_{(T^\dagger)^{ij}T^{ij},k}-iG^{12}_{(T^\dagger)^{ij}T^{ij},k} = \\
=& \frac{1}{4}\Int{k'}\Int{p}\Int{q}\Int{r}\Int{s} \delta_{k+p-q}\delta_{k'+r-s}D^{ij}_{p,q}D^{ij}_{r,s}\left(\avr{\kelphi{p}{2}\kelphi{-q}{2}\kelphi{-r}{1}\kelphi{s}{1}}-\avr{\kelphi{p}{1}\kelphi{-q}{1}\kelphi{-r}{2}\kelphi{s}{2}}\right)=\\
=& \frac{1}{4}\int_{k',p,q,r,s}\!\!\!\!\!\!\!\! \delta_{k+p-q}\delta_{k'+r-s}D^{ij}_{p,q}D^{ij}_{r,s}\left[ \delta_{p-q}\delta_{r-s}\left(iG^{22}_piG^{11}_r-iG^{11}_piG^{22}_r\right) +\left(\delta_{p-r}\delta_{q-s}+\delta_{p+s}\delta_{q+r}\right)\left(iG^{12}_piG^{21}_q-iG^{21}_piG^{12}_q\right)\right]= \\
=& \frac{1}{4}\Int{p}\left((D^{ij}_{p,p+k})^2+D^{ij}_{p,p+k}D^{ij}_{p+k,p}\right)\rho_p\rho_{p+k}(n_p-n_{p+k}). \label{rhoTT}
\end{align}
Now we take $\ve{k}=0$ and expand the first factor of the integral kernel in Eq.~(\ref{rhoTT}):
\begin{align}
\left. (D^{ij}_{p,p+k})^2\right|_{\ve{k}=0} =& \left. D^{ij}_{p,p+k}D^{ij}_{p+k,p}\right|_{\ve{k}=0} =\left[\frac{2p^ip^j}{\omega^2-2\omega\tilde{\omega}}( K_{\tilde{\omega}+\omega,\ve{p}}- K_{\tilde{\omega},\ve{p}})\right]^2 \overset{\omega\rightarrow 0}{\approx} \left(\frac{p^ip^j}{\tilde{\omega}}\frac{\partial K_{\tilde{\omega},\ve{p}}}{\partial\tilde{\omega}}\right)^2 +\mathcal{O}(\omega). \label{expandDsqu}
\end{align}
The linear term of $\rho_{T^\dagger T}$ in $\omega$ in the long-wavelength limit is the shear viscosity $\eta$. Using Eq.~(\ref{expandDsqu}) and also expanding the spectral function $\rho$ and the thermal factor $~{n_p-n_{p+k}}$ up to first-order in $\omega$ we get: 
\begin{align}
\eta =& \lim\limits_{\omega\rightarrow 0}\frac{\rho_{(T^\dagger)^{12}T^{12}}(\omega,\ve{k}=0)}{\omega} = \lim\limits_{\omega\rightarrow 0}\frac{1}{2\omega}\Int{p}\left[\left(\frac{p^1p^2}{\tilde{\omega}}\frac{\partial K_{\tilde{\omega},\ve{p}}}{\partial\tilde{\omega}}\right)^2 +\mathcal{O}(\omega)\right]\left[\rho_{\tilde{\omega},\ve{p}}^2+\omega\rho_{\tilde{\omega},\ve{p}}\frac{\partial\rho_{\tilde{\omega},\ve{p}}}{\partial\tilde{\omega}}+\mathcal{O}(\omega^2)\right]\left(-\omega \frac{\partial n_{\tilde{\omega}}}{\partial\tilde{\omega}}+\mathcal{O}(\omega^2)\right) = \\
=& \frac{1}{2}\Int{p}\left(\frac{p^1p^2}{\tilde{\omega}}\frac{\partial K_p}{\partial\tilde{\omega}}\rho_p\right)^2(-n'_{\tilde{\omega}}).
\end{align}

\section{Scalar source term}\label{sourceTerm}
To explore the effect of non-zero vacuum-expectation value of $\varphi$, we make the identification $\varphi=\xi+\phi$ in the formulae of appendices \ref{propagators}, \ref{enmomTensor}, \ref{visco} and handle $\phi$ as a classical field, i.e.\ without Keldysh-indices. We wish to prescribe the condition $\avr{\varphi}=\phi$. Substituting $~{\varphi=\xi+\phi}$ into the action with source field $J$ we get
\begin{align}
S[\varphi] &= \frac{1}{2}\Int{x}\Int{y}\varphi_x^\dagger\mathcal{K}_{x-y}\varphi_y +\underbrace{\frac{1}{2}\Int{x}\left(\varphi_x^\dagger J_x+\varphi_xJ_x^\dagger\right)}_{=:S_J[\varphi]}= \frac{1}{2}\Int{p}\varphi_{-p}\varphi_pK_p +\frac{1}{2}\Int{p}\left(\varphi_{-p}J_p+\varphi_pJ_{-p}\right) \\
&\overset{\varphi=\xi+\phi}{=} \frac{1}{2}\Int{p}\xi_{-p}\xi_pK_p +\frac{1}{2}\Int{p}\left(\phi_{-p}J_p+\phi_pJ_{-p}\right)+ \frac{1}{2}\Int{p}\underbrace{\left(\xi_{-p}\phi_pK_p+\phi_{-p}\xi_pK_p+\xi_{-p}J_p+\xi_pJ_{-p}\right)}_{\overset{!}{=}0}.
\end{align}
The elimination of the $\xi$-linear terms imposes the constraint $K_p\phi_p=-J_p$. The energy-momentum tensor has an additional term coming from ${~S_J[\varphi]}$. Collecting the terms according to the field-combinations $\xi\xi$, $\phi\phi$ and $\xi\phi$ we arrive at
\begin{align}
T^{\mu\nu}_k &=\frac{1}{2}\Int{p}\Int{q}\delta_{k+p-q}D^{\mu\nu}_{p,q}(\xi_{-p}+\phi_{-p})(\xi_q+\phi_q) -\Int{p}\Int{q}\delta_{k+p-q}\frac{p^\mu(p+q)^\nu}{q^2-p^2}J_q(\xi_{-p}+\phi_{-p}) \\
&=\Int{p}\Int{q}\delta_{k+p-q}\left(\mathcal{D}^{\mu\nu}_{p,q}\xi_{-p}\xi_q +\mathcal{E}^{\mu\nu}_{p,q}\phi_{-p}\phi_q +\mathcal{F}^{\mu\nu}_{p,q}\xi_{-p}\phi_q \right), \label{enMomWsource}
\end{align}
\end{widetext}
where the corresponding kernel functions read as
\begin{align}
\mathcal{D}^{\mu\nu}_{p,q} &= \frac{1}{2}D^{\mu\nu}_{p,q}, &\\
\mathcal{E}^{\mu\nu}_{p,q} &= \frac{1}{2}D^{\mu\nu}_{p,q}+\frac{p^\mu(p+q)^\nu}{q^2-p^2}K_q, &\\
\mathcal{F}^{\mu\nu}_{p,q} &= \frac{D^{\mu\nu}_{p,q}+D^{\mu\nu}_{-q,-p}}{2}+\frac{p^\mu(p+q)^\nu}{q^2-p^2}K_q.
\end{align}
Only the terms proportional to $\xi\xi$ and $\phi\phi$ contribute to the average $\avr{T^{\mu\nu}}$, resulting an extra term compared to the case of $\phi\equiv 0$. Now we choose a spatially homogeneous and temperature dependent background as follows:
\begin{align}
\phi_p=\delta_p\sqrt{\frac{B(T)}{K_{p=0}}}.
\end{align}
After this, we are left with
\begin{align}
\Int{k}\avr{T^{\mu\nu}_k} &= \avr{T^{\mu\nu}_{0,k}} +B,
\end{align}
which leads to exactly the results we mentioned in Sec.~\ref{thermoConsist}: 
\begin{align}
\varepsilon=\varepsilon_{\phi\equiv 0} +B,&\,\, &P=P_{\phi\equiv 0}-B.
\end{align}
For calculating the spectral function $\rho_{T^\dagger T}$, first we re-observe Eq.~(\ref{enMomWsource}). The expectation value of those terms containing odd number of $\xi$ or $\phi$ fields vanishes. Terms with only $\phi$ fields cancel each other in the anti-commutator, since those do not carry Keldysh-indices. Writing out the remaining ones explicitly: \\
\begin{widetext}
\begin{align}
iG^{21}_{T^\dagger T} = \Int{p}\Int{q}\Int{r}\Int{s}\delta_{k+p-q}\delta_{k'+r-s} &\left(  \mathcal{D}^{\mu\nu}_{p,q}\mathcal{D}^{\mu\nu}_{r,s} \avr{\xi_p^2\xi_{-q}^2\xi_{-r}^1\xi_s^1}+ \mathcal{F}^{\mu\nu}_{p,q}\mathcal{F}^{\mu\nu}_{r,s}\avr{\xi_p^2\xi_{-r}^1}\phi_{-q}\phi_s +\right.\\
 &\left. +\mathcal{D}^{\mu\nu}_{p,q}\mathcal{E}^{\mu\nu}_{r,s}\avr{\xi_p^2\xi_{-q}^2}\phi_{-r}\phi_s +\mathcal{D}^{\mu\nu}_{r,s}\mathcal{E}^{\mu\nu}_{p,q}\phi_p\phi_{-q}\avr{\xi_{-r}^1\xi_s^1}\right), \\
iG^{12}_{T^\dagger T} = \Int{p}\Int{q}\Int{r}\Int{s}\delta_{k+p-q}\delta_{k'+r-s} &\left(  \mathcal{D}^{\mu\nu}_{p,q}\mathcal{D}^{\mu\nu}_{r,s} \avr{\xi_p^1\xi_{-q}^1\xi_{-r}^2\xi_s^2}+ \mathcal{F}^{\mu\nu}_{p,q}\mathcal{F}^{\mu\nu}_{r,s}\avr{\xi_p^1\xi_{-r}^2}\phi_{-q}\phi_s +\right.\\
 &\left. +\mathcal{D}^{\mu\nu}_{p,q}\mathcal{E}^{\mu\nu}_{r,s}\avr{\xi_p^1\xi_{-q}^1}\phi_{-r}\phi_s +\mathcal{D}^{\mu\nu}_{r,s}\mathcal{E}^{\mu\nu}_{p,q}\phi_p\phi_{-q}\avr{\xi_{-r}^2\xi_s^2}\right).
\end{align}
The spectral function of the composite operator ${~(T^\dagger)^{\mu\nu}T^{\mu\nu}}$ is the difference of the two above-written formulae:
\begin{align}
\rho_{T^\dagger T} &=iG^{21}_{T^\dagger T}-iG^{12}_{T^\dagger T} = \rho_{T^\dagger T,\,0}+ \\
& +\Int{p}\Int{q}\Int{r}\Int{s}\delta_{k+p-q}\delta_{k'+r-s}\mathcal{F}^{\mu\nu}_{p,q}\mathcal{F}^{\mu\nu}_{r,s}\phi_{-q}\phi_s \delta_{p-r}\rho_p + \label{rhoTTSourcePlusTerm1}\\
&+\Int{p}\Int{q}\Int{r}\Int{s}\delta_{k+p-q}\delta_{k'+r-s}\left(\mathcal{D}^{\mu\nu}_{r,s}\mathcal{E}^{\mu\nu}_{p,q}\phi_p\phi_{-q}\delta_{r-s}(iG^{11}_r-iG^{22}_r)+ \mathcal{D}^{\mu\nu}_{p,q}\mathcal{E}^{\mu\nu}_{r,s}\phi_{-r}\phi_s\delta_{p-q}(iG^{22}_p-iG^{11}_p)\right). \label{rhoTTSourcePlusTerm2}
\end{align}
\end{widetext}
The first additional term compared to the case of $~{\phi\equiv 0}$ is Eq.~(\ref{rhoTTSourcePlusTerm1}). In case of homogeneous background ($~{\phi_p\sim\delta_p}$) it simplifies to
\begin{align}
\sim (\mathcal{F}^{\mu\nu}_{k,0})^2\rho_k \overset{\ve{k}=0}{\rightarrow} 0,
\end{align}
since for space-space indices all the three kernel functions vanish in the long-wavelength limit $~{p=0}$, if either of their arguments vanishes: 
\[\begin{array}{ccccc}
\left.\mathcal{D}^{ij}_{p,q=0}\right|_{\ve{p}=0}=0,& \,\, & \left.\mathcal{E}^{ij}_{p,q=0}\right|_{\ve{p}=0}=0, &\,\, &\left.\mathcal{F}^{ij}_{p,q=0}\right|_{\ve{p}=0}=0.
\end{array}\]
\begin{widetext}
Eq.~(\ref{rhoTTSourcePlusTerm2}) is the second additional term to $\rho_{T^\dagger T,\,0}$. For homogeneous, non-zero background it reads as
\begin{align}
\sim & \delta_k\mathcal{E}_{0,0}^{\mu\nu}\left(\Int{r}\mathcal{D}_{r,r}^{\mu\nu}(iG^{11}_r-iG^{22}_r) + \Int{p}\mathcal{D}_{p,p}^{\mu\nu}(iG^{22}_p-iG^{11}_p)\right)=0, \nonumber
\end{align}
which vanishes for any $\mu$ and $\nu$ pairs. Whereas neither Eq.~(\ref{rhoTTSourcePlusTerm1}) nor Eq.~(\ref{rhoTTSourcePlusTerm2}) contribute to the spectral function ${~\rho_{T^\dagger T}}$, the expression of the shear viscosity does not modify in the case of a homogeneous, temperature dependent background.
\end{widetext}

\bibliographystyle{apsrev4-1}
\raggedright
\bibliography{etaOspaper}
\end{document}